\documentclass[a4paper,twoside]{article}
\usepackage{qic,epsfig}

\usepackage{amsmath}
\usepackage{cite}
\usepackage{MissingT}

\renewcommand{\includegraphics}[2][\relax]
             { \ifx\relax#1
                 \epsfig{file=#2.eps}
               \else
                 \epsfig{file=#2.eps,#1}
               \fi
               \vspace{1.0em}
             }
\renewcommand{\caption}[1]{\fcaption{#1}}

\begin{document}
\setlength{\textheight}{8.0truein}    

\runninghead{Graphical algorithms and
             threshold error rates for the 2d colour code}
            {D. S. Wang, A. G. Fowler, C. D. Hill, L. C. L. Hollenberg}

\normalsize\textlineskip
\thispagestyle{empty}
\vspace*{0.88truein}
\setcounter{page}{1}
\alphfootnote
\fpage{1}

\centerline{\bf GRAPHICAL ALGORITHMS AND}
\centerline{\bf THRESHOLD ERROR RATES FOR THE 2D COLOUR CODE}
\vspace*{0.37truein}
\centerline{\footnotesize
        D. S. WANG\footnote{dswang@physics.unimelb.edu.au} ,
        A. G. FOWLER, C. D. HILL, L. C. L. HOLLENBERG}
\vspace*{0.015truein}
\centerline{\footnotesize\it Centre for Quantum Computer Technology,}
\baselineskip=10pt
\centerline{\footnotesize\it School of Physics, University of Melbourne,}
\baselineskip=10pt
\centerline{\footnotesize\it Victoria 3010, Australia}
\vspace*{0.21truein}

\begin{abstract}
Recent work on fault-tolerant quantum computation making use of
topological error correction shows great potential, with the 2d
surface code possessing a threshold error rate approaching $1\%$
\cite{Raus07d, quant-ph:0905.0531}.  However, the 2d surface code
requires the use of a complex state distillation procedure to achieve
universal quantum computation.  The colour code of \cite{Bomb06} is a
related scheme partially solving the problem, providing a means to
perform all Clifford group gates transversally.  We review the colour
code and its error correcting methodology, discussing one approximate
technique based on graph matching.  We derive an analytic lower bound
to the threshold error rate of $6.25\%$ under error-free syndrome
extraction, while numerical simulations indicate it may be as high as
$13.3\%$.  Inclusion of faulty syndrome extraction circuits drops the
threshold to approximately $0.1\%$.
\end{abstract}



\section{Introduction}
\label{sec:introduction}

The development of quantum error correcting codes in 1995
\cite{Shor95, Cald95, Stea96} is a major milestone in the journey
towards realising a quantum computer that is able to outperform
classical computers for large problems.  Error correction allows the
suppression of decoherence rate during a quantum algorithm, allowing
one to perform lengthy calculations such as Shor's algorithm for prime
number factorisation \cite{Shor94b} with high fidelity results.  The
threshold theorem \cite{Knil96b} states that, provided all gates are
constructed with a failure rate below some threshold error rate,
arbitrary length quantum computation can be achieved by employing
quantum error correction with polylogarithmic overhead.

The act of concatenation, the recursive grouping of logical qubits
into successively higher level logical qubits, is one method to form
codes with a threshold.  However, this concatenation procedure creates
non-local stabilisers involving an ever increasing number of physical
qubits.  As such, threshold error rates for codes formed in this
manner suffer when one is limited to local interactions in few
dimensions.  For example, the $7$-qubit Steane code has a threshold of
$p_{\textrm{th}} = 1.85 \times 10^{-5}$ \cite{Svor06} when restricted
to a 2d lattice with only nearest-neighbour couplings, and the
Bacon-Shor code performs similarly,
$p_{\textrm{th}} = 2.02 \times 10^{-5}$ \cite{spedalieri2008llt}.
On the other hand, topological error correcting codes are designed
with such locality constraints in mind and hence are particularly well
adapted to these architectures.  It has been shown that the 2d surface
code \cite{Kita97b} possesses a threshold error rate approaching $1\%$
\cite{Raus07d, quant-ph:0905.0531}.  Additionally, use of defect
braiding permits for long-range, multi-qubit interactions
\cite{Raus06, Fowl08}.

The major drawback of the 2d surface code lies in its use of state
distillation in performing $S$ and $T$ gates to achieve universal
quantum computation.  This method requires one be able to produce mass
produce logical qubits approximating the states

\begin{equation}
  \ket{Y} = (\ket{0} + i \ket{1})/\sqrt{2},
  \qquad
  \ket{A} = (\ket{0} + e^{i\pi/4} \ket{1})/\sqrt{2}.
\end{equation}

\noindent Several approximations to one of these states are run
through a distillation circuit to produce a single state better
approximating that state.  For example, the $\ket{Y}$ state
distillation circuit (corresponding to implementing $S$ gates) takes
seven states approximating $\ket{Y}$ as input to produce a better
approximate to $\ket{Y}$.  Through several levels of iteration, one
can arrive at the desired state with arbitrary accuracy.  This
procedure requires a large number of qubits, potentially several
orders of magnitude greater than the rest of the computer, dedicated
to producing such states.

This motivates work towards finding a topological scheme which
bypasses the issue of state distillation.  In this paper we will
consider the \emph{colour code} of \cite{Bomb06}, which may be adapted
to incorporate the desirable features of the surface code, whilst
preserving its ability to implement $S$ gates transversely
\cite{quant-ph:0806.4827v1}.  Although the issue of state distillation
remains for $T$ gates, it is significant progress towards a
distillation free topological code.  At present what is missing is a
determination of the error threshold of the colour code.

Due to the similarity between the surface code and the colour code,
and the relative sizes of their stabilisers, one can argue that the
threshold for the colour code should be between $10^{-4}$ and
$10^{-3}$, but this has yet to be demonstrated numerically.  Recently,
the threshold for a different colour code scheme featuring a honeycomb
lattice has been found to be $p_{\textrm{th}} = 0.109(2)$ by mapping
to a random $3$-body Ising model \cite{cond-mat:0902.4845}.  In these
topological codes, one typically makes heavy use of classical
computation in order to diagnose the sources of errors from a given
syndrome.  This task is non-trivial and one must have some efficient
algorithm when applying these codes in real life.  In order to
determine the threshold for such a code, one in fact makes use of
exactly this procedure, for it is necessary to restore the state back
into a codeword and hence determine the the logical failure.

Following our work on the surface code, we devise a method that may be
applied in real life without modification to correct errors on the
colour code based on finding approximate hypergraph matching
solutions.  Using this approach, we find the average time to failure
of an encoded quantum memory under both error-free (\emph{ideal}) and
error-prone (\emph{non-ideal}) syndrome extraction.  We also determine
the threshold analytically under ideal syndrome extraction by
combinatoric arguments, without the burden of performing the
hypergraph matching.  We find the threshold for the 2d colour code
under ideal syndrome extraction to be lower bounded by
$p_{\textrm{th}} \geq 6.25\%$, and numerical simulations indicate it
may be as high as $p_{\textrm{th}} = 13.3\%$ comparable to the
honeycomb colour code \cite{cond-mat:0902.4845} and the surface code
\cite{Denn02, quant-ph:0905.0531} under similar circumstances.
However, inclusion of the syndrome extraction circuits drops the
threshold error rate to approximately $0.1\%$.

This paper is organised as follows.  Section \ref{sec:error
correction} reviews the colour code and briefly discusses how error
correction is performed, assuming that some logical state has been
encoded into the surface.  The discussion of logical qubits and
logical operations is deferred until section \ref{sec:logical qubits}.
We return to the details of error correction in section
\ref{sec:hypergraph mimicry}.  The simulation procedure and the
results under non-ideal syndrome extraction are presented in section
\ref{sec:results}.  Section \ref{sec:ideal results} presents threshold
estimates using two different methods under ideal syndrome extraction,
firstly by simulation and then by combinatorics.

\section{Error correction on the 2d colour code}
\label{sec:error correction}

\begin{figure}
\centering
  \begin{minipage}{0.05\linewidth}
  ~
  \end{minipage}
  \begin{minipage}{0.55\linewidth}
    \includegraphics{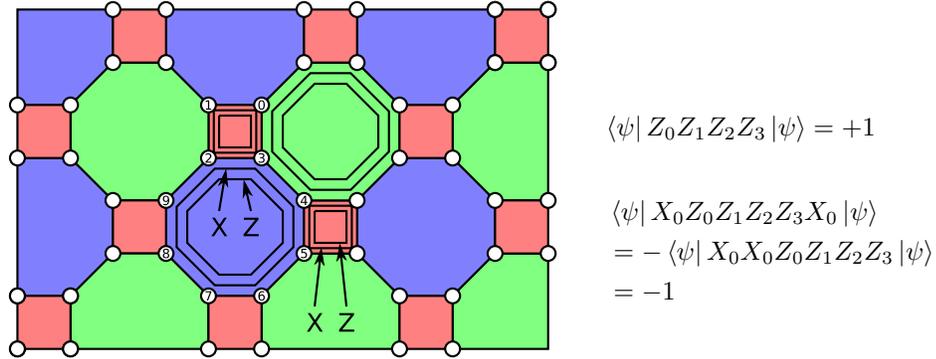}
  \end{minipage}
  \begin{minipage}{0.2\linewidth}
    \begin{align*}
      \bra{\psi} Z_0 Z_1 Z_2 Z_3 \ket{\psi} &= +1
    \end{align*}
    \begin{align*}
    & \bra{\psi} X_0 Z_0 Z_1 Z_2 Z_3 X_0 \ket{\psi} \\
    &   = -\bra{\psi} X_0 X_0 Z_0 Z_1 Z_2 Z_3 \ket{\psi} \\
    &   = -1
    \end{align*}
  \end{minipage}
  \begin{minipage}{0.05\linewidth}
  ~
  \end{minipage}

\caption{2d lattice of qubits for the colour code.  White circles
represent data qubits.  Ancilla qubits located within the plaquettes
are not shown.  The stabiliser generators are the tensor products of
$X$ and $Z$ on the qubits around each plaquette:
$X_0 X_1 X_2 X_3$,
$Z_0 Z_1 Z_2 Z_3$,
$X_2 X_3 X_4 X_5 X_6 X_7 X_8 X_9$,
$Z_2 Z_3 Z_4 Z_5 Z_6 Z_7 Z_8 Z_9$, etc.  The state $\ket{\psi}$ is
initialised to the simultaneous $+1$ eigenstate of all the generators.
A single $X$ error, $\ket{\psi} \rightarrow X \ket{\psi}$, will be
observed as an eigenvalue change on the adjacent $Z$-stabilisers due
to the commutation relation between the Pauli matrices.}

\label{fig:stabilisers}
\end{figure}

Consider the 2d lattice of qubits arranged as shown in figure
\ref{fig:stabilisers}, assuming for now that the lattice extends
indefinitely.  Each plaquette is associated with two generators of the
stabiliser group; the tensor product of the Pauli-$X$ matrix, $X$, on
the qubits around its perimeter, and the tensor product of the
Pauli-$Z$ matrix, $Z$, on those same qubits.  Neighbouring plaquettes
always share two qubits, ensuring all stabilisers commute.  Assigned
to each stabiliser is a colour, red, green or blue, such that each
qubit belongs to exactly one $X$-stabiliser and one $Z$-stabiliser of
each colour.  The plaquette colours shown in figure
\ref{fig:stabilisers} are not inherent to the system, merely a device
to aid error correction.  For the purposes of this paper, red
stabilisers will always be synonymous with square stabilisers. 

\begin{figure}
\centering
  \includegraphics{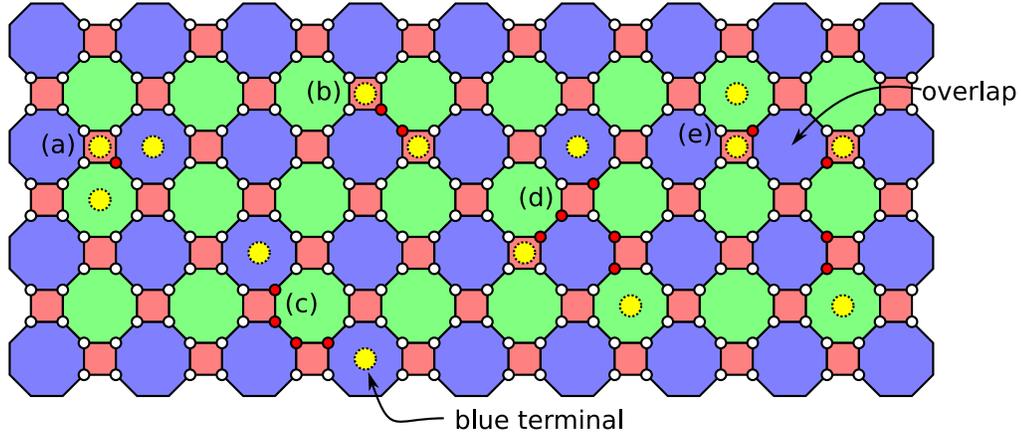}

\caption{Examples of syndromes produced by error chains (colour
online).  A red circle indicates a physical error on the given data
qubit.  A yellow circle at a site indicates a change in eigenvalue
from its previous measurement.
(a) a single error toggles three plaquette eigenvalues.
(b) $2$-chain generating two red terminals.
(c) $2$-chain generating two blue terminals.
(d) $3$-chain generates three different colour terminals.
(e) many-terminal error chain can be decomposed into $2$-chains and
$3$-chains with overlapping terminals.}

\label{fig:syndrome}
\end{figure}

In the absence of errors, the state of the system is the simultaneous
$+1$ eigenstate of each stabiliser.  An $X$-error on one qubit causes
the state to toggle between the $\pm 1$ eigenvalue states of the
adjacent $Z$-stabilisers due to the commutation relations.  Ancilla
qubits located within each plaquette allow the eigenvalue to be
measured locally.  The configuration of eigenvalue changes measured
forms the \emph{syndrome}, from which the physical location of the
errors can be inferred.  Similarly, $Z$-errors are detected by
measuring the eigenvalues of the $X$-stabilisers.  Figure
\ref{fig:syndrome} illustrates more complex possible syndromes.

Combinations of errors often conspire in such a way to conceal the
intermediate eigenvalue flips of many plaquettes.  These sets of
errors, or \emph{error chains}, may be seen as the primitives
generating the observed syndrome instead of the independent errors on
individual qubits.  We will also regard single errors as error chains.
The eigenvalue changes an error chain induces are its
\emph{terminals}.  The colour of a terminal is the colour of the
plaquette whose eigenvalue it alters.  Error chains may have two same
colour terminals ($2$-chains), three terminals with one of each colour
($3$-chains), or more terminals.  Any chain having in excess of three
terminals may be decomposed into superpositions of lower order chains,
with some terminals overlapped (figure \ref{fig:syndrome}e).  Indeed
any $2$-chain (figure \ref{fig:syndrome}b) may be derived from a pair
of $3$-chains, however they are treated as a primitive due to their
relative simplicity.  Since all observable syndromes are generated by
superpositions of error chains, the identification of error chains is
equivalent to locating errors.  Given an arbitrary syndrome, error
chain identification is carried out using algorithms from graph theory
as follows.

A \emph{matching}, $M$, of an undirected graph, $G$, is a subgraph of
$G$ such that each node in $M$ has exactly one incident edge.  A
\emph{perfect matching} is a matching where all nodes in $G$ belong to
$M$.  A minimum-weight perfect matching is an element from the set of
perfect matchings, whose sum of edge weights is minimised.  Many
minimum-weight perfect matchings may be possible, in which case we
recover just one.  Polynomial-time matching algorithms for graphs
exist, for example Edmonds' blossom algorithm \cite{edmonds1965pta,
edmonds1965mma, Cook99}.  Unfortunately, error correction in the
colour code requires a hypergraph matching algorithm, for which
efficient algorithms are not known.  A \emph{hypergraph} is a
generalisation of a graph, where edges are promoted to
\emph{hyperedges}, sets of arbitrary numbers of vertices.  The
\emph{rank} of the hypergraph is the maximum cardinality hyperedge.
This will be discussed further in section
\ref{sec:hypergraph mimicry}, where our implemention of the hypergraph
matching is detailed.  For now, we assume that such an algorithm
exists.

Given a syndrome, the eigenvalue changes observed translate to nodes
on a hypergraph.  Any possible $2$-chain required to generate a pair
of terminals is represented by an edge joining the corresponding pair
of nodes.  Similarly, a hyperedge is added for every possible
$3$-chain.  A matching on this hypergraph then represents a
corresponding set of error chains which together will partially
generate the syndrome, because each terminal belongs to exactly one
error chain.  Thus a perfect matching will reproduce the entire
syndrome, allowing error correction to be performed.

\begin{figure}
\centering
  \includegraphics{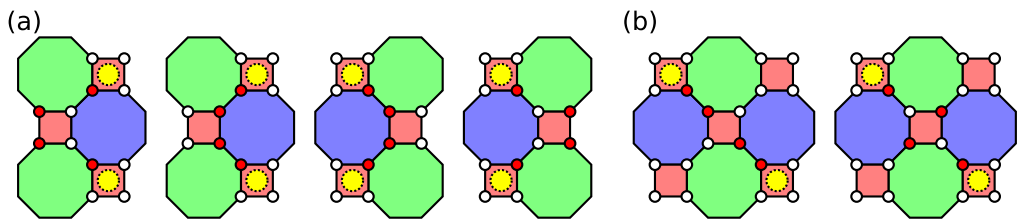}

\caption{Two different syndromes caused by four errors.  Case (a) is
more probable than case (b) as it can be generated by more length-$4$
error chains, and one should weight it accordingly in the graph to
recover the matching of maximum likelihood.  Our simulations do not
consider such fine details, instead we choose only one of the
possibilities covering each case.  Incorporating these alternatives
could potentially increase the threshold.}

\label{fig:degenerate}
\end{figure}

Many matchings can be found for a given syndrome.  Since the syndrome
arises from physical errors which have a low probability of occurring,
one should rig the weights of the edges and hyperedges such that the
matching algorithm finds the matching of maximum likelihood.  In
general, the weight of an edge between some terminals should take into
consideration all of the different possible error chains generating
those terminals.  For example, although the terminals in both cases
shown in figure \ref{fig:degenerate} are equidistant, figure
\ref{fig:degenerate}a should be weighted as more probable than figure
\ref{fig:degenerate}b purely because there are more length-$4$ error
chains generating the former setup.  However, in order to simplify
matters, we will not take this into consideration; when many error
chains are possible, we choose the graph edge to represent only one of
the minimum-length possibilities.  Under this approximation, the
weight of an edge is taken to be the error chain length (which is
proportional the logarithm of the probability), so that a chain formed
by 2 errors is weighted identically to two independent single error
chains.  Although a minimum-weight perfect matching under this
approximation will \emph{not} necessarily be the matching of maximum
likelihood, it will nevertheless reproduce the observed syndrome using
the fewest number of errors; the hypergraph is constructed in a way
that the weight-sum of its minimum-weight matching is the minimum
number of errors required to reproduce the syndrome under ideal
syndrome extraction.

The question immediately arises as to what happens when one corrects
along a path of qubits which did not suffer errors.  The discussion
requires a formal introduction to logical qubits and logical gates, so
we only make some brief remarks.  The distance of a code, $d$, is
defined to be the minimum number of physical operations that must be
applied to the physical qubits encoding some logical state in order to
interchange between the two logical states without generating any
observable syndrome.  Under ideal syndrome extraction, a code can in
principle correct any $\floor{\frac{d-1}{2}}$ error events.  This
holds true for the colour code when one corrects by minimum-weight
hypergraph matching.  If fewer than $\floor{\frac{d-1}{2}}$ errors
occur, error correction will always produce rings of operators instead
of chains of operators connecting boundaries.  The rings of operators
are stabilisers of the code, thus the logical state is restored.

As described, many-terminal error chains such as in figure
\ref{fig:syndrome}e are not handled, resulting in misidentification of
the causes of syndromes; this particular syndrome can be produced by
$4$ errors, however it currently appears to the error corrector as a
$10$ error event (generated by a pair of $2$-chains).  One solution is
to add these error chains into the hypergraph in the form of higher
cardinality hyperedges.  However, this not only increases the number
of hyperedges exponentially, but it also increases the complexity of
the matching algorithm.  There is a more elegant solution which rests
on the fact that the plaquette eigenvalue observed determines only the
parity, not the exact number of terminals, at that location; one can
artificially insert a pair of \emph{dummy nodes} into the graph for a
plaquette suspected of harboring overlapping terminals, exactly as if
two terminals had been observed there.  Misplaced dummy pairs can in
the worst case be matched to one another by a weight-$0$ edge, whence
matching continues as if the pair had not been introduced.  Our
simulations need not anticipate such instances: as we have access to
the number of toggles at each plaquette, we introduce a pair whenever
an eigenvalue changes twice or more.  However, in a real
implementation such information is not available so one should devise
some clever algorithm to identify these overlaps and introduce these
nodes only as required.  The benefit from introducing dummy pairs is
that one eliminates the need to match arbitrary rank hypergraphs (only
rank-$3$ is necessary).  Note that even the introduction of dummy
pairs on every plaquette still incurs only a polynomial-time overhead.

\begin{figure}
\centering
  \includegraphics{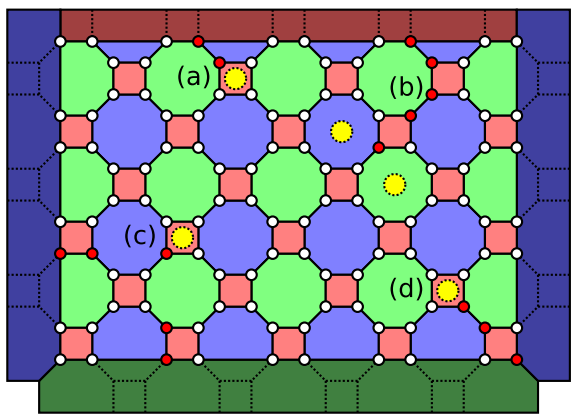}

\caption{Examples of syndromes on finite lattices.  Dark plaquettes
show the different colour boundaries, where eigenvalues are not
measured.
(a) $2$-chain with masked red terminal.
(b) $3$-chain with masked red terminal.
(c) $3$-chain with two masked terminals.
(d) the interface between a green and a blue boundary can also be
considered a red boundary.}

\label{fig:syndrome with boundaries}
\end{figure}

A physical implementation of this code must be done on a finite
lattice, and hence the presence of boundaries which can hide otherwise
visible terminals (figure \ref{fig:syndrome with boundaries}).  There
are three boundary colours.  A red boundary is the interface shielding
red terminals from being observed, so that, for example, a lone red
node can be produced by a $2$-chain from a red boundary.  Green and
blue boundaries are defined equivalently.  Note that the interface
between green and blue boundaries can also be considered a red
boundary.  The case of a $3$-chain producing exactly two mixed-colour
terminals and one masked terminal is easily accommodated for by adding
an edge between those two terminals in the hypergraph, with the
implicit prescription that this edge denotes generating these two
terminals by joining them to the closest boundary by a $3$-chain.
Indeed, any edge or hyperedge between some arbitrary number of nodes
can denote whatever means is necessary to generate exactly those
terminals, independently of all other terminals observed.

The single terminal case is more involved.  Let $G$ denote the
hypergraph one constructs as described so far.  The intent is to
create beside each node, $A$, an associated \emph{boundary node},
$A'$.  Each node is joined to its own boundary node, corresponding to
generating that terminal independently, for example by a $2$-chain to
the closest same colour boundary.  However, this change by itself
permits only one perfect matching: because the degree of every
boundary node is exactly $1$, every node must be matched to its
boundary.  If two regular nodes, $A$ and $B$, are matched together,
their respective boundary nodes, $A'$ and $B'$, are unmatched and thus
this is not a valid perfect matching.  The resolution is to create a
subhypergraph within the boundary nodes mirroring $G$ with only
weight-$0$ edges and hyperedges, so that when $A$ is matched to $B$
(or $B$ and $C$ by a hyperedge), then $A'$ is readily matched to $B'$
($B'$ and $C'$) at no extra cost.  While it is not true that the
boundary node's matching will always reflect that of the regular
node's, this is of no concern; error correction works with only edges
and hyperedges in the matching involving at least one regular node.
This extra change successfully deceives the matching algorithm into
behaving as desired, and error correction on finite lattices may be
performed.

\begin{figure}
\centering
  \includegraphics{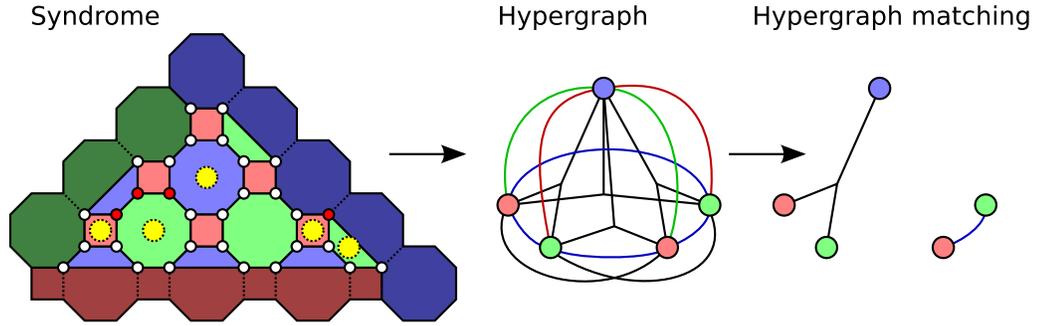}

\caption{The hypergraph constructed from a the observed syndrome.
Edge weights and boundary nodes have not been included. An edge
between two different coloured terminals is coloured only to serve as
a reminder that it denotes a $3$-chain to the other coloured boundary.
The hypergraph matching identifies a corresponding set of error chains
which, once corrected, restores the state of the system into a
codeword state.}

\label{fig:hypergraph matching}
\end{figure}

Figure \ref{fig:hypergraph matching} illustrates the error correction
methodology described here.  It is not strictly limited to this colour
code, and may be adapted to other similarly oriented 2d topological
schemes, such as the honeycomb colour code.  Indeed, the error
correction methodology of the 2d surface code is a specialisation of
the same technique, whereby one forms and matches only rank-$2$
hypergraphs (ie. graphs).

\section{Logical qubits and logical gates}
\label{sec:logical qubits}

\begin{figure}
\centering
  \includegraphics[width=0.8\linewidth]{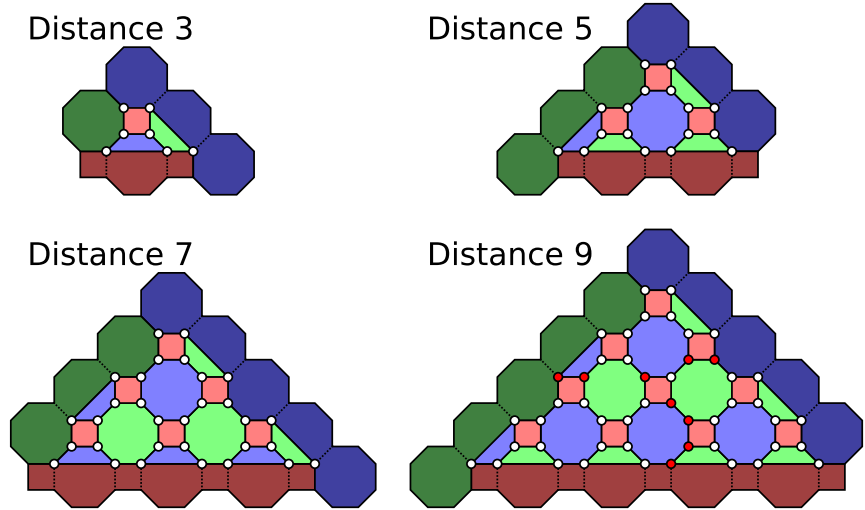}

\caption{Triangular lattices permitting Clifford group gates to be
performed transversely.  Shown are distance $3$, $5$, $7$ and $9$
logical qubits, and the different coloured boundaries in dark.  Also
shown on the distance-$9$ lattice is an example of a logical-$X$ (or
logical-$Z$) operation; a completely masked $3$-chain.}

\label{fig:distances}
\end{figure}

So far we have described the error correction procedure on the colour
code, which preserves some quantum state encoded onto the surface.  In
order to perform computation, logical qubits must be introduced.
Following the original paper \cite{Bomb06}, each logical qubit is
encoded onto a triangular lattice, as shown in figure
\ref{fig:distances}.

The logical-$X$ operation is defined to be any $3$-chain of
$X$-operations on the data qubits connecting together all three colour
boundaries, or any $2$-chain from a qubit along one boundary to the
opposite same coloured boundary.  The symmetry between $X$ and
$Z$-stabilisers constrains the logical-$Z$ operation to be the same
$3$-chains but with $Z$ applied to the sites.  Furthermore, these
chains must be odd in length to yield the correct commutation
relations.  The minimum length of these chains defines the
\emph{distance} of the code.  Therefore all distances $d$ will be
assumed to be odd hereafter.  Note that none of these operations
change the syndrome because all terminals are masked by boundaries.

The nature of the $X$ and $Z$ stabilisers allows logical Hadamard
operations to be performed transversely; when applying the Hadmard
gate on every qubit, the identity $Z = HXH$ implies that $X$-chains
will transform to $Z$-chains, and vice-versa.

The primary interest in using the colour code is its ability to
perform transversal $S$ gates.  Since each plaquette comprises four or
eight qubits, and neighbouring plaquettes share an even number of
qubits, when an $S$ gate is applied to every qubit, every $\ket{0_L}$
state acquires the same phase, and similarly for every $\ket{1_L}$
state.  Furthermore, for all odd distance codes, an odd number of
qubits is enclosed within the triangle lattice, ensuring that
$\ket{0_L} \rightarrow \ket{0_L}$ and
$\ket{1_L} \rightarrow \pm i \ket{1_L}$.

Finally, controlled-not gates may be implemented transversely between
two different sheets of triangular lattices; this operation will
propagate $X$-chains from control qubit to target qubit, and
$Z$-chains from target to control.

\begin{figure}
\centering
  \includegraphics{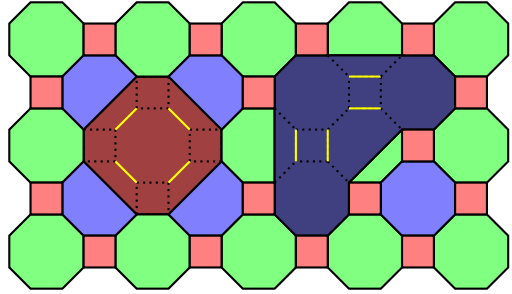}

\caption{Examples of defects on the colour code.  Green and blue
defects share the same form.  Logical qubits are formed by one defect
of each colour, allowing the colour code to adopt many of the benefits
of the 2d surface code, with the additional benefit of transversal $S$
gates \cite{quant-ph:0806.4827v1}.}

\label{fig:defects}
\end{figure}

It is possible to adapt the colour code so that the entire quantum
computer shares a single code \cite{quant-ph:0806.4827v1}, creating
logical qubits by introducing defects into the surface --- contiguous
regions where the eigenvalues of particular stabiliser generators are
no longer measured, thus providing extra degrees of freedom --- in a
similar vein to surface codes \cite{Fowl08b}.  In this manner, it is
possible to recover many of the features of surface codes such as long
range gates.  In addition, one can isolate the triangular lattices
above using a defect of each colour, examples of which are shown in
figure \ref{fig:defects}, thus permitting transversal $S$ gates.  The
details go beyond the scope of this paper.

\section{Hypergraph mimicry and pair assignment}
\label{sec:hypergraph mimicry}

An essential requirement for the colour code error correction
procedure as formulated in section \ref{sec:error correction} is the
existance of an efficient rank-$3$ hypergraph weighted perfect
matching algorithm.  Whether an efficient algorithm exists is unclear;
additional information from, for example, the specialised structure
and the geometry may assist the problem.  Recovering the
minimum-weight hypergraph matching, which presumably produces better
results than matchings of higher weight-sum, is not strictly necessary
and may be relaxed to achieve an efficient error corrector.

While we have formulated the error correction problem such that any
syndrome can be represented as a hypergraph matching, let us clarify
that we never match the hypergraph directly; they are strictly limited
to discussion.  Our procedure is to reduce the initial hypergraph
problem down to a simpler but approximate graph matching problem which
we can solve.  A solution to the graph problem may be mapped back to
give a hypergraph matching and thus is a candidate for error
correction, although it will not necessarily have as low a weight-sum
as the minimum-weight hypergraph matching.

\begin{figure}
\centering
  \includegraphics[width=0.8\linewidth]{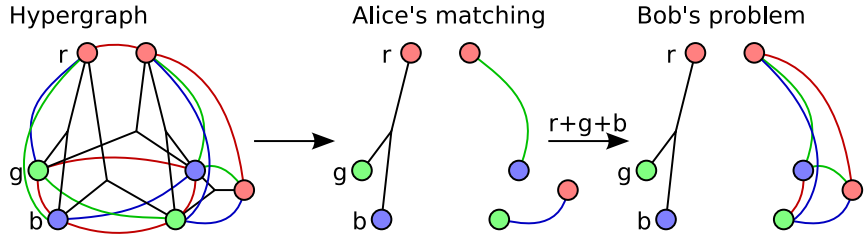}

\caption{Alice informs Bob exactly which nodes match as hyperedges.
Using this information, Bob factors out those nodes from the
hypergraph.  Since he also knows the remaining nodes are matched by
edges, he also eliminates the hyperedges, reducing the hypergraph down
to a graph which can be matched in polynomial time.}

\label{fig:alice full}
\end{figure}

Consider the following scenario: Alice and Bob each have the
hypergraph ahead of time, to which Alice has found the minimum-weight
matching with weight-sum $W_{\textrm{hyper}}$.  Alice wishes to
communicate enough information to Bob, such that Bob can still
reproduce her result in polynomial time.  She can opt to send Bob all
of the hyperedges in her matching.  Bob, knowing the remaining nodes
are matched by edges, can remove all hyperedges from the initial
hypergraph to form a graph, which he can match efficiently to recover
the remainder of Alice's matching (figure \ref{fig:alice full}).

It is possible for Alice to communicate only two of the three nodes in
each matched hyperedge, and still have Bob recover her minimum-weight
hypergraph matching efficiently.  To do so, Bob must collapse the
indicated pairs into a single node.  For each of Alice's matched
hyperedges, Bob replaces the two nodes indicated by a single node, and
the hyperedges connecting these two nodes in the original hypergraph
are replaced by edges.  Because he knows no other nodes are matched by
hyperedges, any remaining hyperedges can be removed and thus Bob's
problem reduces again to matching a graph (figure \ref{fig:alice
partial}).

\begin{figure}
\centering
  \includegraphics[width=0.8\linewidth]{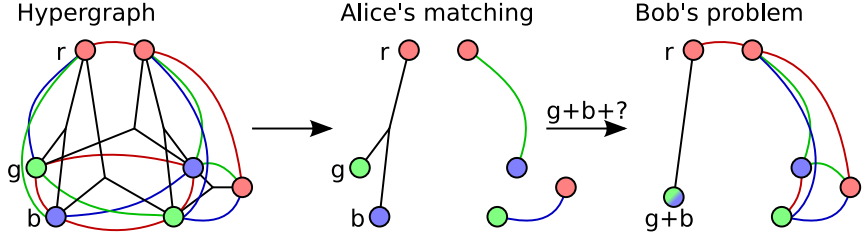}

\caption{Alice informs Bob which green and blue nodes together form
hyperedges.  In order for Bob to recover Alice's matching, he
collapses the indicated blue and green nodes into a single node, so
that hyperedges involving both blue and green nodes become edges, and
all other edges are discarded.  Since Bob also knows that all other
nodes do not form triplets, the remaining hyperedges may also be
discarded.  Thus Bob's problem reduces to a graph matching problem.}

\label{fig:alice partial}
\end{figure}

Obviously we do not have the luxury of Alice's input, and so in both
scenarios illustrated we would instead perform an exhaustive search
over all of Alice's possible inputs.  Such a search exhibits
exponential behaviour which may be counteracted by introducing
approximations; trying only a small subset of the potential inputs at
the expense of finding the true minimum-weight hypergraph matching.
In order to scale up this approximation to higher distance codes, we
will devise a speculative algorithm following Alice's second technique
for it has a smaller domain.

\begin{figure}
\centering
  \includegraphics{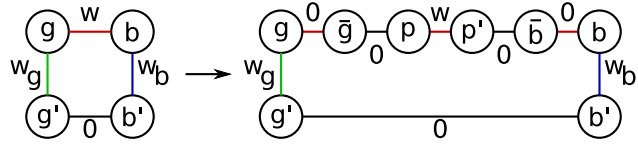}

\caption{Recall that edges connecting green and blue nodes in the
hypergraph implies a connection to a red boundary by a $3$-chain.
Construction of the mimic graph begins by replacing each of these
edges with a series of edges, and introducing four new nodes:
$\bar{g}$, $p$, $p'$, $\bar{b}$.  The newly introduced $p$ act as $g$
and $b$ combined, and $p'$ its boundary.  The edge between $p$ and
$p'$ carries the weight of the original edge.  The $\bar{g}$ and
$\bar{b}$ nodes are necessary ensure only $g$ and $b$ are not used in
two error chains; once independently and once via the $p$ node.}

\label{fig:mimic}
\end{figure}

Our method to assign together pairs of nodes is also achieved by
matching a specialised \emph{mimic graph}, which we now construct.
Let us temporarily neglect all red nodes and build the hypergraph
formed by just the green and blue nodes.  Since we have not included
red nodes, in reality this is simply a graph.  In order to account for
the $O(n_r n_g n_b)$ rank-$3$ hyperedges without actually introducing
hyperedges, one must insert $O(n_g n_b)$ extra nodes.  This can be
done by substituting each edge connecting green and blue nodes with a
series of edges, via four newly introduced intermediate nodes:
$\bar{g}$, $p$, $p'$, $\bar{b}$ (figure \ref{fig:mimic}).

\begin{figure}
\centering
  \includegraphics{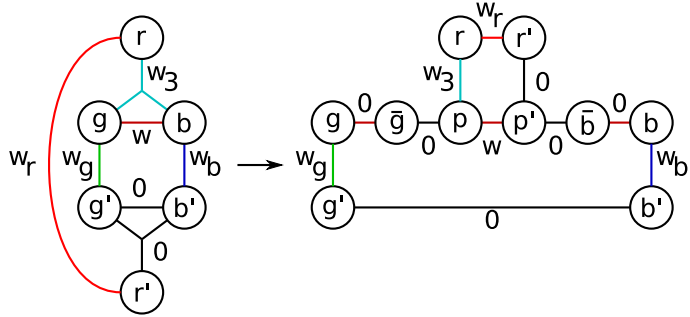}

\caption{Red nodes in the hypergraph are incorporated into the mimic
graph.  The hyperedge between $r$, $g$, $b$ in the original hypergraph
becomes, an edge between $r$ and $p$ in the mimic graph.  The red
boundary node, $r'$, is also joined to the pair boundary node, $p'$ by
a weight-$0$ edge.}

\label{fig:mimic with red}
\end{figure}

This transformation in itself does not affect matchings on this graph;
the choice of matching the edge ${g,b}$ becomes a choice of matching
an alternating set of edges.  However, we have the new interpretation
of $p$ being the amalgam of $g$ and $b$, and $p'$ its boundary.  The
special nodes $\bar{g}$ and $\bar{b}$ are \emph{disables}, which
always have degree $2$.  They serve to ensure each node does not
participate in two error chains, once as an individual and once as a
pair.  The interpretation of $p$ as the pair formed by combining $g$
and $b$ allows us to add the hyperedges into the mimic graph; edges
joining red nodes to pair nodes.  As before, we join the respective
boundaries together to fool the perfect matching condition (figure
\ref{fig:mimic with red}).

We still need to incorporate the case of generating red and green
terminals by connecting to a blue boundary, and similarly for the red
and blue terminals connecting to the closest green boundary.  We can
join red nodes to green nodes (and their respective boundary nodes)
with the implicit assumption that those two terminals join to the
closest blue boundary to form a $3$-chain.  However, for our
approximate method it is best to minimise the degree of the nodes.  We
instead choose to insert additional blue nodes along the blue boundary
plaquettes, used solely for forming these $3$-chains.  Whether or not
the boundary actually masked a terminal is unimportant; an introduced
node can at worst be matched to its own boundary node by a weight-$0$
edge and matching continues as if they had not been introduced, akin
to the introduction of dummy pairs.  Similarly, we introduce green
nodes along the green boundary to account for joining red and blue
terminals by a $3$-chain.  The procedure incurs a polynomial overhead
and can be optimised to minimuse the number of extra nodes.

\begin{figure}
\centering
  \includegraphics{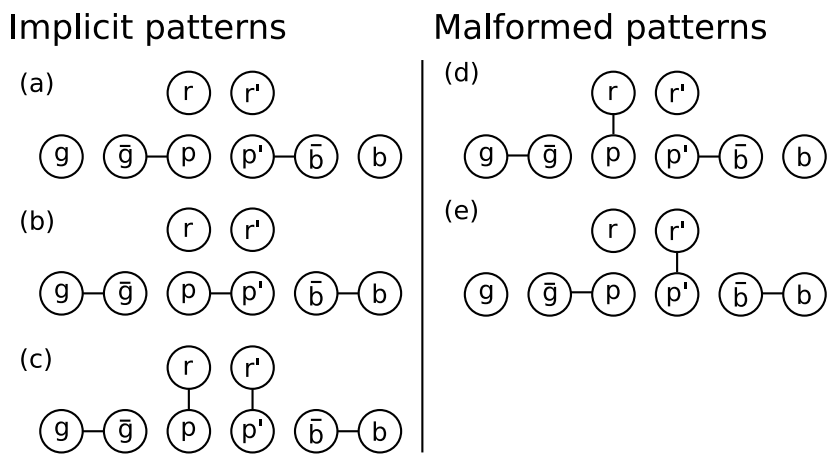}

\caption{Left shows the three (partial) patterns implicitly demanded
in the mimic matching if it is to always be interpretable as a
correction.  Malformed patterns may arise in the mimic matching,
leaving this interpretation invalid.
(a) extra nodes introduced were unused.
(b) green and blue connect together to closest red boundary.
(c) red, green and blue form $3$-chain.}

\label{fig:implicit patterns}
\end{figure}

This mimic graph emulates some of the properties of the hypergraph and
can be matched efficiently.  In particular, all matchings on the
hypergraph are matchings on the mimic graph.  However, the converse if
\emph{not} true; matchings of mimic graphs may not necessarily be
translated into hypergraph matchings.  This is due to an implicit
demand for one of three particular patterns within the triplet region
(figure \ref{fig:implicit patterns}), when including the red nodes.
The matching algorithm knows nothing of such patterns, nor does the
placement of weights assist it.  Thus what we extract from the mimic
graph matching itself is \emph{not} a correction in general; only when
it does not contain any malformed patterns is this true.  However, we
can use the matching as a choice for pair assignment to then produce a
correction.  We identify the matching of $r$ to $p$ and the subsequent
matching of $g$ to $\bar{g}$ (due to $\bar g$ having degree-$2$) as
the assignment of $r$ to $g$.

\begin{figure}
\centering
  \includegraphics{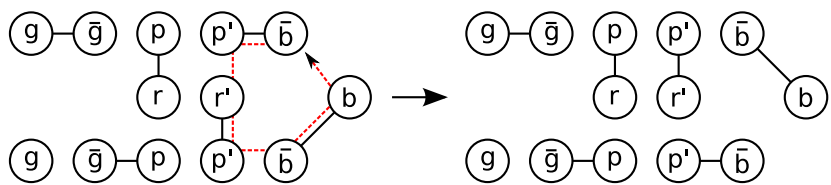}

\caption{Malformed matchings can often be corrected by adding a
weight-$0$ alternating cycle, shown in dashed lines.  Adding an
alternating cycle to a matching toggles the edges between included and
excluded from the matching, yielding a new matching.}

\label{fig:rotation}
\end{figure}

There are two reasons for this choice.  First, such combinations will
always arise if a hypergraph matching were translated into a mimic
graph matching.  Second, one can often rotate the right hand side by a
weight-$0$ alternating cycle, forming the desired patterns (figure
\ref{fig:rotation}).  An alternating cycle on a graph is simple cycle
with edges alternating between included and excluded from the
matching.  The weight of an alternating cycle is the sum of the
weights of edges not in the matching minus the weight of edges in the
matching, so that the new matching will have weight $W+w$.  One could
correct the mimic matching by searching for lowly weighted alternating
cycles, though this can be inefficient as one must be careful about
breaking existing well-formed patterns.  We observe that for these
simple cycles, one can instead rematch with $r$ and $g$ collapsed into
a single node, thus avoiding these complicated searches.

From this choice of pair assignments, we collapse the rank-$3$
hypergraph with dummy nodes down to a \emph{trial graph}, as Bob had
done previously after receiving Alice's partial input.  This trial
graph may be matched efficiently to give a trial solution with some
weight-sum $W_{\textrm{trial}}$.  A single choice of assignment can go
amiss, so we take the solution with the lowest weight-sum over several
initial assignments.  In addition, there are six simple mimic graph
constructions possible --- one for each permutation of red, green and
blue\footnote{
For the mimic graph permutation we have worked with, we can crudely
identify the red-to-pair edge in the mimic graph as the $3$-chain.
The mimic matching may not be a correction as it allows for blue
terminals to be either not corrected, corrected once, or corrected
twice (by a $2$-chain and a $3$-chain).  Red and green terminals do
not suffer such problems; they will always be corrected exactly once.
Thus even the interchange of green and blue can produce different
results.
} --- and, in general, the minimum-weight matching of each can have a
different weight-sum.  The variants with the \emph{highest} weight-sum
matching, $W_{\textrm{mimic}}$, are taken to yield matchings that
closest approximate the minimum-weight hypergraph matching, and hence
give better choices of pair assignment and trial solutions.  Only the
matchings from these variants are used as choices of pair assignment.
This probabilistic method appears to yield near-optimal matchings even
as the codeword distance increases.  The reason we assume that higher
weight-sum mimic graph matchings give better trial solutions is due to
the relative ordering of the weight-sums:
\begin{equation}
  W_{\textrm{mimic}} \leq W_{\textrm{hyper}} \leq W_{\textrm{trial}}
\end{equation}

In general, we do not know the weight-sum of the minimum-weight
rank-$3$ hypergraph matching, $W_{\textrm{hyper}}$.  However, as we
have constructed the mimic graph to encompass possible hypergraph
matchings, the mimic matching weight-sum is upper bounded by the
minimum-weight hypergraph matching weight-sum.  Extra freedom from the
mimic matching not necessarily being a correction allows for its
weight-sum to be lower: $W_{\textrm{mimic}} \leq W_{\textrm{hyper}}$.
Similarly, after deciding upon a choice of assignment, many potential
hypergraph matchings are discarded.  Thus all trial matchings must
have weight-sum $W_{\textrm{trial}} \geq W_{\textrm{hyper}}$.  Notice
that if $W_{\textrm{trial}} = W_{\textrm{mimic}}$, the heuristic has
not introduced any additional errors over the minimum-weight
hypergraph matching; the trial matching itself is a minimum-weight
hypergraph matching.

It is the number of uncertain matchings we use to gauge the quality of
the approximation.  We find that using up to $6 \times 25$ initial
trials, this method has a $95\%$ probability of recovering the
minimum-weight hypergraph matching with certainty from a single time
step with $\frac{d+1}{2}$ errors scattered for large lattices.  Taking
$6 \times 50$ trials does not significantly increase the probability.
From the remaining uncertain cases, the final correction applied
typically has a weight-sum only one greater than the mimic matching
weight-sum, so that if it were not the hypermatching it at least falls
very close.  For our simulations, we have chosen $6 \times 25$ initial
trials.

\section{Simulation results}
\label{sec:results}

Simulations take place on the triangular lattices of figure
\ref{fig:distances}.  In our simulations we determine the average
number of syndrome extraction cycles a quantum state encoded in a
colour code endures before error correction results in a logical
failure.  The simulations trace only the propagation of $X$-errors
throughout the machine.

A single simulation instance proceeds as follows.  First, the quantum
computer is initialised perfectly in the simultaneous $+1$ eigenstate
of every stabiliser generator.  At each timestep, each qubit has a
probability $p$ of a memory error, then the syndrome information is
extracted simultaneously over the entire surface.  Red syndrome is
extracted by preparing the ancilla positioned within each red
stabiliser in the $\ket{0}$ state ($\ket{+}$ state for $Z$-syndromes).
Subsequently four controlled-nots are directed inwards (outwards) from
the surrounding data qubits to the ancilla, which are then measured in
the $Z$-basis ($X$-basis).  A coloured plaquette syndrome is extracted
by first preparing a $4$-qubit cat-state.  Each qubit in the cat-state
interacts with two distinct data qubits in that stabiliser, and is
measured independently.  The four measurements together give the
parity at that stabiliser.  Further details of syndrome extraction are
available in \cite{quant-ph:0806.4827v1}.

A memory error on a qubit is a probability $p/3$ of incurring either
an $X$, $Z$ or $Y = XZ$ error.  Preparation of $\ket{0}$ and $\ket{+}$
states each have a probability $p$ of resulting in the $\ket{1}$ and
$\ket{-}$ states respectively.  Similarly, measurement in the $X$ and
$Z$-bases have a probability $p$ of obtaining the incorrect result.
Finally, the two-qubit gates have an equal probability $p/15$ of each
of the $15$ non-trivial tensor products of $I$, $X$, $Y$ and $Z$.

\begin{figure}
\centering
  \includegraphics[width=0.8\linewidth]{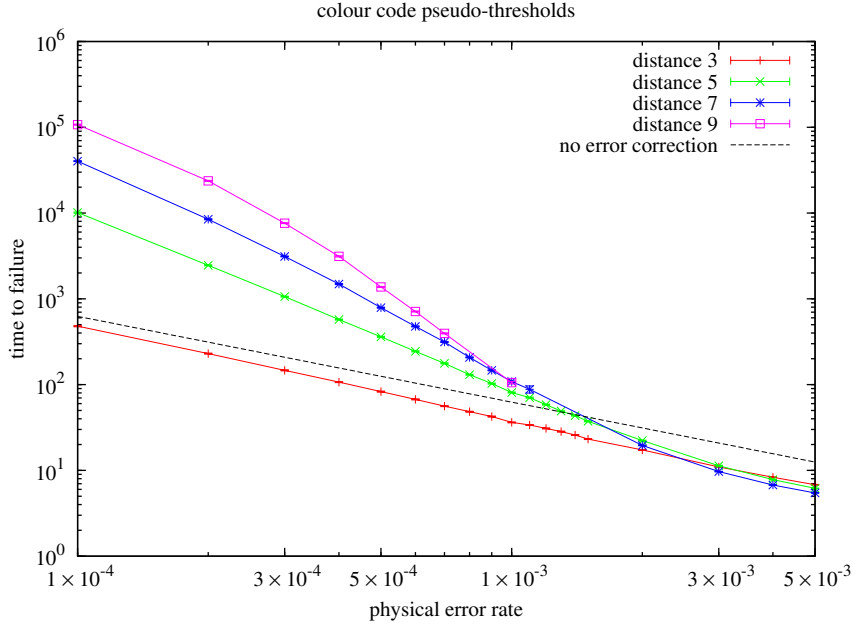}

\caption{The average time to failure of a quantum memory in the colour
code.  The vertical axis is proportional to real time.  Error bars
represent the uncertainty in the average time taken over a large
number of instances.  The asymptotic threshold error rate is
approximately $0.1\%$.}

\label{fig:faulty result}
\end{figure}

After each timestep, the syndrome information is used to correct the
state and check for logical failure.  Errors during syndrome
extraction are taken into account by collating syndrome information
over time, forming a 3d structure of eigenvalue changes.  Error chains
are permitted to span through time, with those segments denoting an
error during syndrome extraction in the prior timestep.  In our
simulations, an error chain segment spanning through time is weighted
equally to one of the same length spanning though space.  In addition,
we also collect one final syndrome ideally before error correction and
checking for logical failure.  Should error correction fail using this
augmented syndrome information, we record the failure time.
Otherwise, we continue to the next timestep recollecting this syndrome
non-ideally.
Using this procedure, the average life expectency of a logical state
is shown figure \ref{fig:faulty result}.  Also shown is the average
lifespan of a single non-error-corrected qubit.

There are two features of significance.  Firstly, the gradients of the
curves are observed to converge to the same value for low error rates
for distance-$5$ and above.  As with the surface code, correlated
errors during the syndrome extraction cycle has caused a distance-$d$
code to correct fewer than the expected $E(d) = \floor{\frac{d-1}{2}}$
errors.  Despite the distance-$3$ colour code being identical to the
$7$-qubit Steane code, the topological method of dealing with errors
during syndrome extraction coupled to our rather simplistic weighting
of error chain segments results in the colour code offering no
benefits over unerror-corrected qubits.  A combination of these
correlated errors during syndrome extraction and our approximate error
correction method are responsible for distance $5$, $7$ and $9$ codes
ultimately being able to correct the same number of errors.  However,
higher distance codes can be seen to improve upon lower distances;
there are fewer combinations of two errors causing a distance-$7$ code
to fail than for distance-$5$, and fewer yet again for distance-$9$.
In light of this, we assert that higher distance codes will eventually
be able to correct more errors under the error correction method
presented.

Secondly, the intersections of successive distance codes is highly
mobile.  Furthermore, the intersections move leftwards thus do not
give a lower bound on the threshold.  This in part may be due to the
presence of boundaries of the lattice; our simulations of the toric
code and the surface code show similar trends
\cite{quant-ph:0905.0531}, though admittedly are much less extreme.
We will revisit this issue in section \ref{subsec:dangerous syndrome
coverage}.  Due to computing limitations, we are unable simulate
higher distance codes to observe a convergence.  The existing results
indicate a threshold of approximately $0.1\%$.

\section{Ideal threshold error rate}
\label{sec:ideal results}

As the asymptotic threshold under realistic syndrome extraction is
difficult to find, we seek to determine the threshold under ideal
syndrome extraction.  In this limit, there is no need to consider
error chains spanning through time, making theoretical calculations
much more accessible.  We proceed in two directions: firstly by direct
simulation, and secondly by counting the number of dangerous syndrome
patterns.

\subsection{Direct simulation}
\label{subsec:direct simulation}

We already have the means to simulate the colour code logical failure
rates under ideal syndrome extraction.  One point of note is that
because syndrome information is always correct, we no longer need to
consider error chains spanning through time; error chain segments
spanning through time are now weighted infinitely greater than those
spanning through space.  As such, each timeslice is corrected
independently of all other timeslices.  This implies that the total
number of possible syndromes is finite, thus one can determine the
logical error rate per timestep by summing over all possible dangerous
syndrome combinations:

\begin{equation}
  p_L^{(d)}(p_0)
    = \sum_{k=0}^{Q} A_d(k) p^k (1-p)^{(Q-k)},
      \quad p = \frac 2 3 p_0
\end{equation}

Here $p_0$ is the physical error rate, $Q(d)$ is the number of data
qubits, and $A_d(k)$ is the number of dangerous syndromes resulting
from $k$ errors in the distance-$d$ code.  The factor of $\frac 2 3$
is due to our error model: only two of the three possibilities $X$,
$Y$, $Z$ may lead to logical-$X$ failures.  For later convenience, we
will make the change of variable $k \rightarrow (F+k)$, where
$F(d) = \frac{d+1}{2}$ is the minimum number of errors required for a
distance-$d$ code to fail under true minimum-weight hypergraph
matching and ideal syndrome extraction.  We will always reference a
prefactor $A_d(F+k)$ by its offset $k$ from the expected leading
order term.

\begin{equation}
  p_L^{(d)}(p_0)
    = p^F \sum_{k=-F}^{Q-F} A_d(F+k) p^k (1-p)^{(Q-F)-k},
      \quad p = \frac 2 3 p_0
\label{eq:logical failure rate}
\end{equation}

\begin{figure}
\centering
  \includegraphics[width=0.8\linewidth]{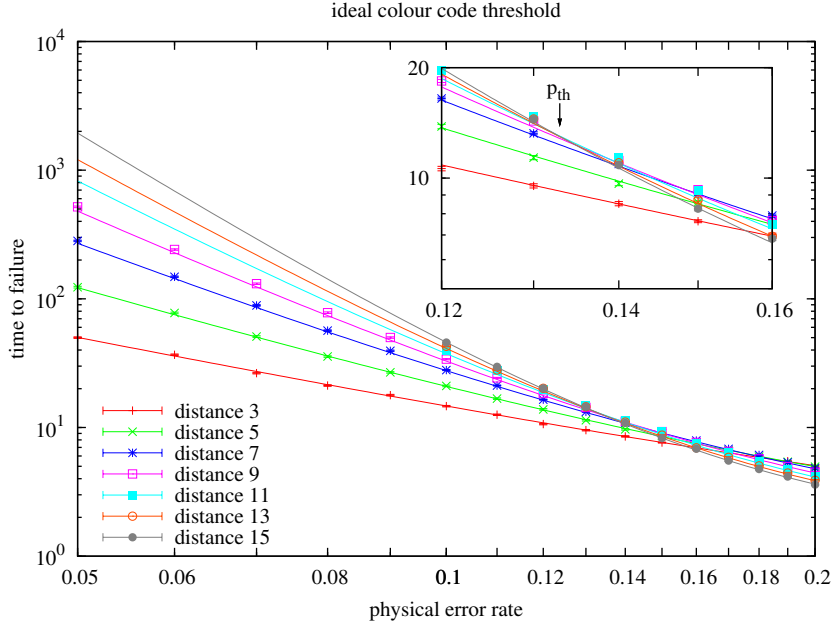}

\caption{Average life expectancy of a quantum state under error free
syndrome extraction.  The minimum error rate at which successive
distance codes intersect is taken to best approximate the asymptotic
threshold, $p_{\mathrm{th}} = 13.3\%$.}

\label{fig:ideal result}
\end{figure}

The exact value for $A_d(F+k)$ depends on the details of the matching
algorithm used, and can be obtained by running the error corrector for
each possibility.  Since the total number of $(F+k)$-error
configurations to test is large, $Q \choose F+k$, we approximate
$A_d(F+k) \approx {Q \choose F+k} r_k$ by determining only the ratio
of $(F+k)$-error failures, $r_k$, from a large random sample.
The results in the ideal syndrome extraction limit by direct
simulation of the error correction procedure are shown in figure
\ref{fig:ideal result}.  The data points are obtained from simulations
of the code at certain error rates, while the curves are given by
equation \ref{eq:logical failure rate}, using our matching algorithm
to estimate the prefactors $A_d(F+k)$.  We observe a threshold of
$p_{\mathrm{th}} = 13.3\%$.

\subsection{Dangerous syndrome coverage}
\label{subsec:dangerous syndrome coverage}

One can place an upper bound on $A_d(F+k)$, the number of dangerous
syndromes as a result of $F+k$ errors, under true minimum-weight
hypergraph matching.  We will assume hypergraph matching hereafter, in
particular with regards to dangerous syndromes, and simplify refer to
it as the \emph{matching}.  By construction of the hypergraph, the
matching will correct for all $E(d) = \frac{d-1}{2}$ error cases:
$A_d(F+k) = 0, \forall~k < 0$.  Non-trivial contributions to the
logical failure rate start from $F(d) = \frac{d+1}{2}$.

\begin{equation}
  p_L^{(d)}(p_0)
    = p^F \sum_{k=0}^{Q-F} A_d(F+k) p^k (1-p)^{(Q-F)-k},
      \quad p = \frac 2 3 p_0
      \label{eq:hypergraph error rate}
\end{equation}

For an $F$ error syndrome to cause failure, the $F$ errors must all
fall on a single length-$d$ logical operator, $\hat{O}_d$;
minimum-weight matching would find it preferable to correct the state
by applying corrections on the remaining $E$ qubits in $\hat{O}_d$,
thus performing the logical operation.  Thus an upper bound to
$A_d(F)$ can be determined by enumerating all length-$d$ logical
operators, then choosing $F$ qubits from each.

\begin{figure}
\centering
  \hfill
  \begin{minipage}{0.25\linewidth}
    \begin{tabular}[t]{|c|r|}
      \hline
       d & $A_d(F)$ \hspace{1.0em} \\
      \hline
       3 &         21 \\
       5 &        332 \\
       7 &       5807 \\
       9 &    100,120 \\
      11 &  1,671,668 \\
      13 & 27,227,258 \\
      \hline
    \end{tabular}
  \end{minipage}
  \begin{minipage}{0.6\linewidth}
    \includegraphics[width=0.8\linewidth]{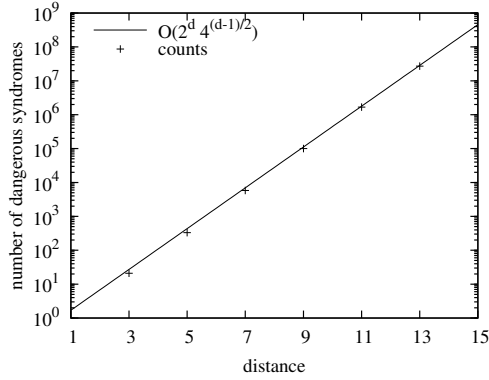}
  \end{minipage}
  \hfill

\caption{The leading contribution to the logical error rate, $A_d(F)$,
when using minimum-weight hypergraph matching grows exponentially with
the distance of the code.  These results were obtained by counting the
length-$d$ logical operators and determining unique combinations of
$F$ errors lying along a single logical operator.}

\label{fig:enum result}
\end{figure}

It is tempting to make the statement that all dangerous syndromes as a
result of $F+k$ errors simply stem from a dangerous $F$-error syndrome
and scattering a further $k$ errors.  Unfortunately, this simplistic
argument is \emph{not} valid, as dangerous syndrome formed otherwise
exist due to the presence of higher length logical operators.  One can
find counter-examples, and the difference in gradient between $A_d(F)$
(figure \ref{fig:enum result}) and $A_d(F+k)$ (figure \ref{fig:general
enum result}) further reinforce their presence.  

\begin{figure}
\centering
  \includegraphics{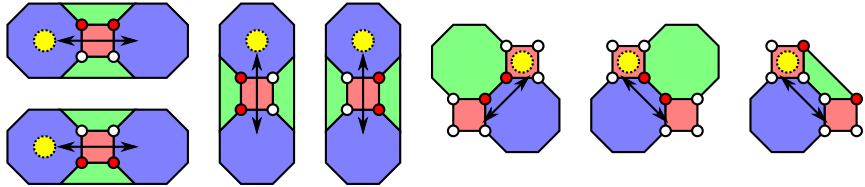}

\caption{Rules of shifting terminals amongst same colour plaquettes.
Each move requires exactly two errors, shown as the red qubits.  Green
and blue terminals share the same rules.  Red terminals have
additional multi-step rules, such as those shown in figure
\ref{fig:degenerate}.  Extra choices may be possible for terminals
beside the boundary.}

\label{fig:shifting rules}
\end{figure}

We conjecture that all dangerous $F+k$ error syndromes are formed by
some combination of $F$ errors on qubits belonging to a single
continuous length-$(d+2\lambda)$ error chain, then scattering a
further $k$ errors onto the remaining qubits, where $\lambda$ is a
constant.  A \emph{continuous} error chain is one which may be derived
placing a single error, generating some terminals, then shifting these
terminals by the rules of figure \ref{fig:shifting rules}.  The rules
are defined such that any logical operator can be formed in this way.
In particular, because length-$(d+2\lambda)$ logical operators are
formed in this way, the conjecture trivially holds true for the
$(F+\lambda)$-error case.  The conjecture is a constraint on the
distribution of errors required for logical failure for a length-$d$
code.  For example, not any arbitrary placement of $F$ errors will
cause a code to fail, only very select combinations, namely those
where the errors occur on qubits belonging to a single length-$d$
logical operator.

One can upper bound the number of continuous length-$(d+2\lambda)$
error chains.  We start by placing an initial error on the $Q$ qubits
on the lattice, generating up to three terminals.  Discarding
backtracking, each terminal can be shifted by one of up to $6$ rules
(figure \ref{fig:shifting rules}).  Some care should be taken when
moving red stabilisers as one has additional options that are not
listed, such as those shown in figure \ref{fig:degenerate}.
Similarly, minor modifications are necessary when a terminal lies next
to a boundary.  Regardless, for a continuous length-$(d+2\lambda)$
error chain, because each shift is accomplished by placing down pairs
of errors, one must make a total of $\frac{d-1}{2} + \lambda$ shifts,
shared between the three terminals.  There are approximately
$\paren{\frac{d-1}{2}}^2$ ways to divides these shifts between the
three terminals.  Thus the maximum number of continuous
length-$(d+2\lambda)$ error chains is:

\begin{equation}
  N_d < Q \paren{\frac{d-1}{2}}^2 6^{(d-1)/2 + \lambda}
\end{equation}

\noindent
Our conjecture then bounds the number of dangerous syndromes as a
result of $F+k$:
\begin{align}
  A_d(F+k)
   &< N_d {d+2\lambda \choose F} {Q-F \choose k}
      \label{eq:crude A_d}
\end{align}

\noindent
Since $F = \frac{d+1}{2}$, it follows that ${d + 2\lambda \choose F}$
is exponentially bounded, using
${\chi \choose \chi/2}
  = \frac{\chi!}{(\chi/2)!(\chi/2)!}
  \leq 2^\chi$:
\begin{align}
  {d + 2\lambda \choose F}
   &\leq
      {d + 2\lambda \choose F+\lambda}
    \leq
      2^{d + 2\lambda}
\end{align}

\noindent
Substituting $A_d(F+k)$ back into equation \ref{eq:hypergraph error
rate}, then simplifying using the binomial expansion:
\begin{align}
  p_L^{(d)}(p_0)
   &< N_d 2^{d+2\lambda} p^F
      \sum_{k=0}^{Q-F} {Q-F \choose k} p^k (1-p)^{(Q-F)-k},
      \quad p = \frac 2 3 p_0                                           \\
   &= N_d 2^{d+2\lambda} p^F
      \label{eq:hypergraph error rate simplified}
\end{align}

For the case of $A_d(F)$, the dangerous $F$-error syndromes are the
various combinations of $F$ errors along the length-$d$ logical
operators.  For a given combination, the continuous error chain
covering it is the logical operator from which it was derived.  In
this case, terminals must always step towards their boundary,
narrowing the choices of shifting terminals down to approximately
$4^{(d-1)/2}$.  We can calculate $A_d(F)$ exactly for hypergraph
matching by first enumerating all length-$d$ logical operators, and
then find all unique combinations of $F$ errors.  The logical
operators themselves can be determined, for example, by deforming some
given initial logical operator by the combinations of the stabiliser
generators.  These results are shown in figure \ref{fig:enum result},
confirming our bound on $A_d(F)$ displays the correct asymptotic
behaviour.

\begin{figure}
\centering
  \hfill
  \begin{minipage}{0.45\linewidth}
    \includegraphics[width=1.0\linewidth]{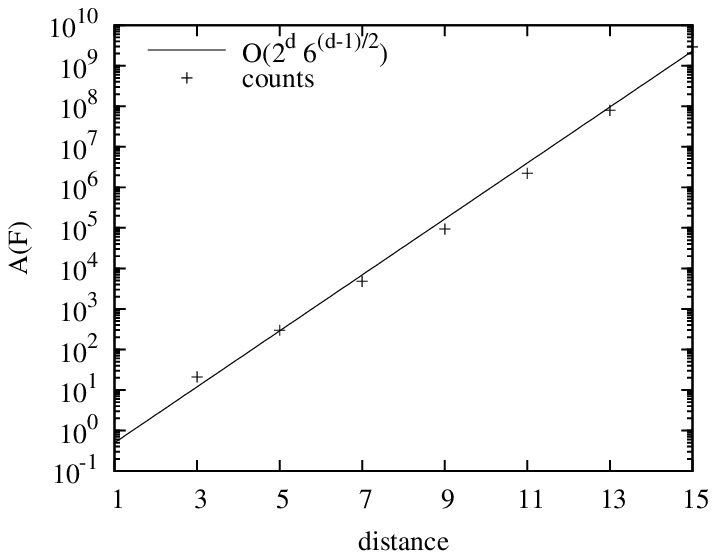}
  \end{minipage}
  \begin{minipage}{0.45\linewidth}
    \includegraphics[width=1.0\linewidth]{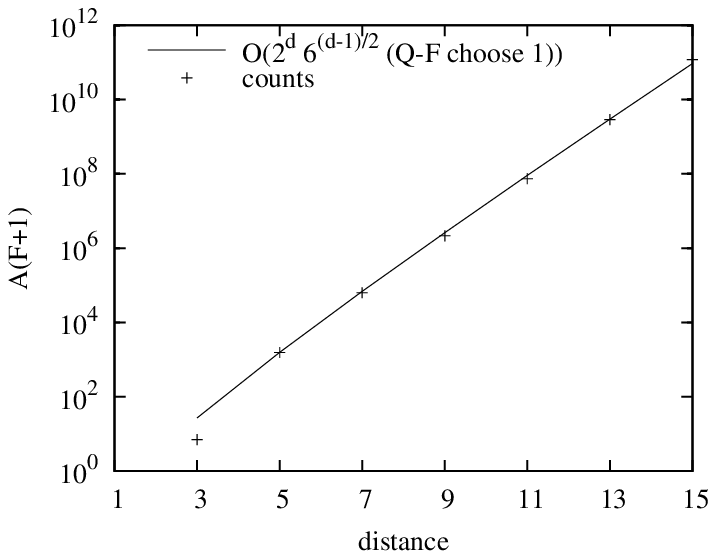}
  \end{minipage}
  \hfill

\vspace{-1.5em}
  \hfill
  \begin{minipage}{0.45\linewidth}
    \includegraphics[width=1.0\linewidth]{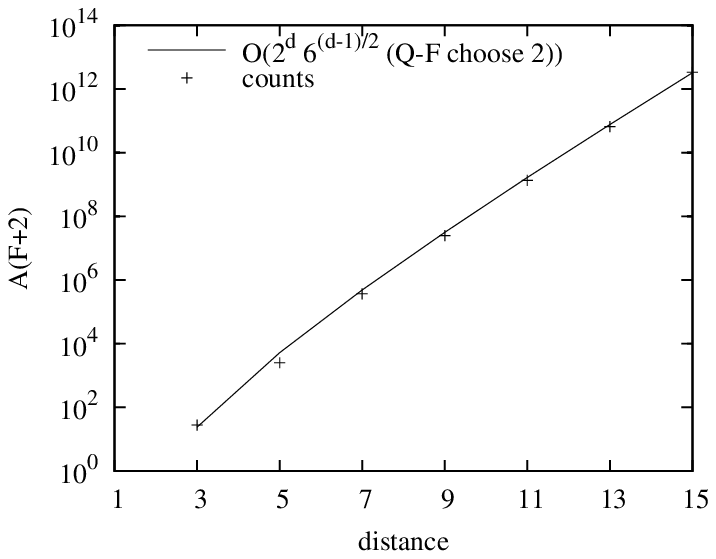}
  \end{minipage}
  \begin{minipage}{0.45\linewidth}
    \includegraphics[width=1.0\linewidth]{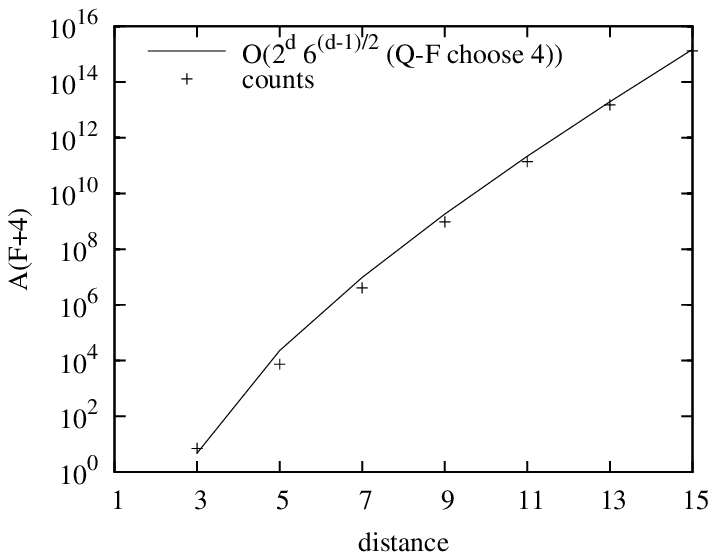}
  \end{minipage}
  \hfill

\vspace{-1.5em}
\caption{The number of dangerous syndromes as a result of $F+k$ errors
is $O\paren{6^{(d-1)/2}~2^d~{Q-F \choose k}}$.  The data points were
obtained using the approximate matching method outlined in section
\ref{sec:hypergraph mimicry}.}

\label{fig:general enum result}
\end{figure}

The higher order prefactors $A_d(F+2)$ and $A_d(F+4)$ are shown
alongside the more general theoretical bound in figure
\ref{fig:general enum result}.  The counts shown in these graphs are
taken from the preceeding simulations since enumerating quickly
becomes difficult.  The reason for choosing $F+2$ is that in the
colour code logical operators have length $d+4n$, should only
contribute to $A(F+2)$ and higher.  Our results show that the
$A_d(F+k)$ gradients agree with the combinatoric error rate, giving
the initial conjecture further merit.  Unfortunately, due to the
approximate nature of our matching, higher length logical operators do
in fact contribute to $A_d(F+1)$ and even $A_d(F)$, so that they too
display the same $6^{(d-1)/2}$ growth rate.

It is not necessary to determine the prefactor in the logical error
rate (equation \ref{eq:hypergraph error rate simplified}) with a great
deal of precision to calculate an asymptotic threshold; the complexity
itself is sufficient.  An asymptotic threshold is obtained by
comparing the logical error rates of successive distance codes in the
limit of large distance, $d$, whereupon all polynomial factors in $d$
disappear:

\begin{align}
  1 &= \lim_{d \rightarrow \infty} \frac{p^{(d+2)}_L}{p^{(d)}_L}    \\
  p_{\textrm{th}}
    &= \frac 3 2 \frac 1 {6 \cdot 2^2} = 6.25\%
\end{align}

\begin{figure}
\centering
  \begin{minipage}{1.0em}
    (a)
  \end{minipage}
  \begin{minipage}{0.45\linewidth}
    \includegraphics[width=\linewidth]{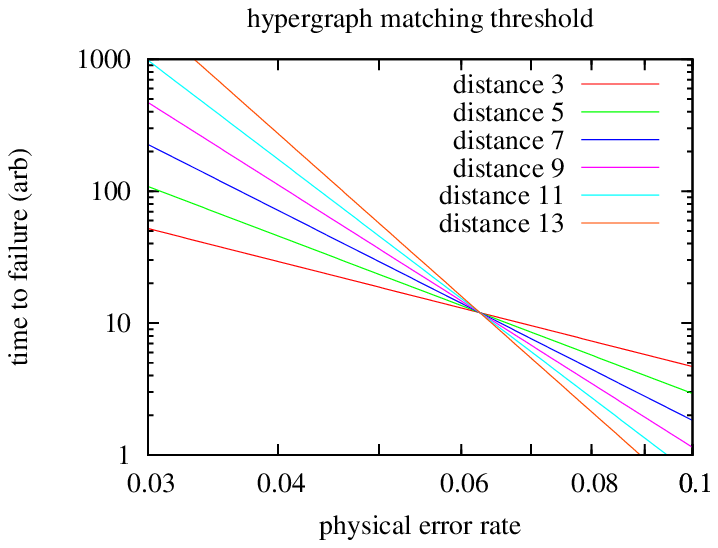}
  \end{minipage}
  \begin{minipage}{1.0em}
    (b) 
  \end{minipage}
  \begin{minipage}{0.45\linewidth}
    \includegraphics[width=\linewidth]{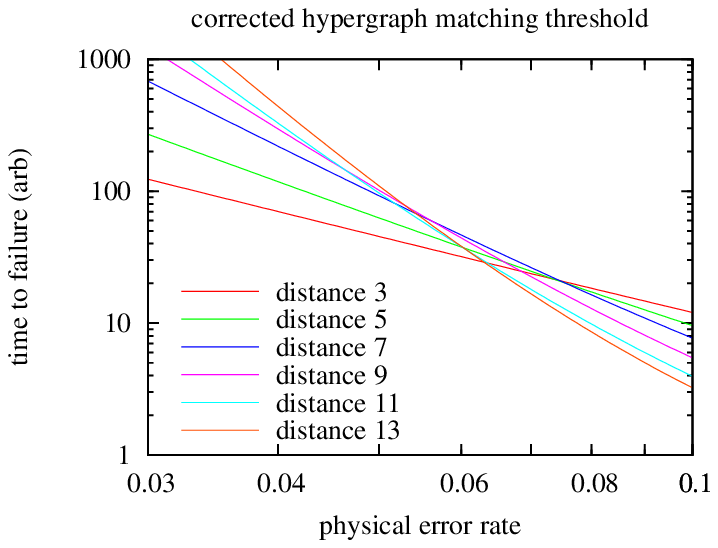}
  \end{minipage}

\vspace{-1.5em}
\caption{
(a) Lower bounds to the expected average time to failure when
correcting by minimum-weight matching under ideal syndrome extraction,
using equation \ref{eq:hypergraph error rate simplified} as the
logical error rate.  The asymptotic threshold is
$p_{\textrm{th}} = 6.25\%$.
(b) The different growth rates between $A_d(F+k)$ for $k \leq 1$ and
$k \geq 2$ can be accounted for by using equation \ref{eq:hypergraph
error rate} to determine the logical error rate.  This leads to
fluctuating pseudo-thresholds.  In this graph, we have also used
$A_d(F+k) = {Q \choose F+k}$ when equation \ref{eq:crude A_d} exceeds
${Q \choose F+k}$.
}

\label{fig:hypergraph result}
\end{figure}

Figure \ref{fig:hypergraph result}a shows the average lifespan of a
quantum memory over different distances using equation
\ref{eq:hypergraph error rate simplified}.  However, we have already
observed that the prefactors $A_d(F+k)$ follow two different growth
rates --- $O(4^{(d-1)/2})$ for $k \leq 1$, and $O(6^{(d-1)/2})$ for
$k \geq 2$ --- hence one should not so readily simplify the equation.
Applying equation \ref{eq:hypergraph error rate} with the additional
clamping of $A_d(F+k) \leq {Q \choose F+k}$ gives figure
\ref{fig:hypergraph result}b.  Interestingly, it features the same
fluctuating pseudo-thresholds as observed in the simulations.  While
we have noted that our simulation $A_d(F)$ grows at the faster rate,
it is presumed that the earlier terms are suppressed at different
rates, giving rise to the same effect.

These results rest on the initial conjecture, from which we have
deduced the growth rates of $A_d(F+k)$.  Simulation results appear to
follow the predicted growth rates.  However, it remains to be
rigorously proven that some constant $\lambda$ exists for all $d$, at
least on this geometry, from which a lower bound to the threshold
evidently follows.

\section{Conclusion}

We have described a general error correction procedure suitable for
many 2d topological codes as a minimum-weight hypergraph perfect
matching problem.  We have also described an efficient but approximate
method for matching rank-$3$ hypergraphs required for this colour
code, which in principle may be used for other $3$-colour codes.  The
method seeks solutions by constructing two graphs: one upper-bounded
by the hypergraph solution, the other lower-bounded.  When the two
bounds meet, we identify that the approximation has introduced no
errors.  Thus this method can in principle be implemented as an
initial pass before less efficient methods, which attempt to improve
the code's performance, for example to ensure that a distance-$d$ code
reliably corrects $\floor{\frac{d-1}{2}}$ errors.

Combinatoric arguments presented here suggest that the asymptotic
threshold error rate of the colour code to be lower bounded by
$p_{\textrm{th}} \geq 6.25\%$ under error free syndrome extraction.
Simulations using the approximate hypergraph matching method show
that the lower bound on the threshold under these conditions may be as
high as $p_{\textrm{th}} = 13.3\%$.  Once faulty syndrome extraction
circuits are introduced, numerical simulations indicate that the
threshold may fall to approximately $0.1\%$.  Unfortunately, this
begins to encroach on the realm of the concatenated codes, which do
not need such complex error correction procedures and some of which
bypass the use of state distillation.  An efficient specialised
matching algorithm for the colour code in the presence of boundaries
may be possible, potentially raising these lower bounds on the
threshold and improving its prospects.


\nonumsection{Acknowledgements}

We thank Ashley Stephens for helpful discussions.
This work was supported by the Australian Research Council, the
Australian Government, and the US National Security Agency (NSA) and
the Army Research Office (ARO) under contract number W911NF-08-1-0527.

\nonumsection{References}
\bibliography{paper}
\bibliographystyle{unsrt}

\end{document}